\documentclass[]{spie}  

\usepackage{amsmath,amsfonts,amssymb}
\usepackage{graphicx}
\usepackage[colorlinks=true, allcolors=blue]{hyperref}
\usepackage{siunitx}
\usepackage{subcaption}

\title{Photonic mid-infrared nulling for exoplanet detection on a planar chalcogenide platform}

\author[a]{Harry-Dean Kenchington Goldsmith}
\author[b]{Michael J. Ireland}
\author[a]{Pan Ma}
\author[a]{Barry Luther-Davies}
\author[c]{Rongping Wang}
\author[d]{Barnaby Noris}
\author[d]{Peter Tuthill}
\author[a]{Stephen J. Madden}
\affil[a]{Laser Physics Center, Research School of Physics and Engineering, Australian National University, Canberra, ACT 2900, Australia}
\affil[b]{Research School of Astronomy and Astrophysics, Australian National University, Canberra, ACT 2611, Australia}
\affil[c]{Research Institute of Advanced Technologies, Ningbo University, Ningbo, 315211, China}
\affil[d]{Sydney Institute for Astronomy (SIfA), School of Physics, University of Sydney, NSW 2006, Australia }

\authorinfo{Further author information: (Send correspondence to H-D.K.G.)\\H-D.K.G.: E-mail: harry-dean.kenchingtongoldsmith@anu.edu.au, Telephone: +61 2 612 50694\\  M.J.I.: E-mail: michael.ireland@cmp.com, Telephone: +61 2 612 50288}

\begin{document} 
\maketitle

\begin{abstract}
The future of exoplanet detection lies in the mid-infrared (MIR). The MIR region contains the blackbody peak of both hot and habitable zone exoplanets, making the contrast between starlight and planet light less extreme. It is also the region where prominent chemical signatures indicative of life exist, such as ozone at 9.7~\si{\um}. At a wavelength of 4~\si{\um} the difference in emission between an Earth-like planet and a star like our own is 80~dB. However a jovian planet, at the same separation exhibits 60~dB of contrast, or only 20~dB if it is hot due to its formation energy or being close to its host star. A two dimensional nulling interferometer, made with chalcogenide glass, has been measured to produce a null of 20~dB, limited by scattered light. Measures to increase the null depth to the theoretical limit of 60~dB are discussed.  
\end{abstract}


\section{INTRODUCTION}
\label{sec:intro}  

Exoplanets are worlds that orbit stars other than our own. They are faint, radiating celestial bodies that have not had a history of fusion - separating them from brown dwarfs. At present 3706~exoplanets (as of 15/03/2018 on NASA exoplanet archive\footnote{https://exoplanetarchive.ipac.caltech.edu/}) have been confirmed, the majority of which used data from the Kepler space telescope via changes to the exoplanet's host star emission. These detections are referred to as indirect detection techniques, with the most common types of techniques being the transit detection method - light being blocked from the host star - and the radial velocity detection method - the exoplanet's gravitational influence inducing a Doppler shift on the host star's light. Few exoplanet detections used the exoplanetary emission itself, or light reflected off the exoplanet. These methods are termed direct detections. Exoplanets where direct detection becomes favourable, or indeed possible, are usually much more massive than Jupiter, often have the same or twice the radius of Jupiter, and are further away from their host star than Jupiter is to our sun~\cite{Marois2008, Lagrange2009, Kalas2009}. 

Large exoplanets far away from their host star are candidates for coronagraphy~\cite{Guyon2005} - where the light of the host star is physically blocked by an occulting spacecraft or by an image-plane mask (coronagraph)~\cite{Guyon2003}. Large star-planet separation assists in this detection as the coronagraph mask blocks any exoplanet too close (typically multiple times Earths's orbit), determined by fabrication tolerances~\cite{Guyon2010}, to the host star. 

Detections from reflected light would theoretically require 60~dB star suppression if they were the size and mass of Jupiter~\cite{Burrows2004}, with a semi major axis within 1~AU. At present there are only two exoplanets detected using reflected light, however these used data from Kepler and did not spatially separate planet from star light~\cite{Charpinet2011}. Natural emissions from the exoplanet may be easier to target. A young giant exoplanet can have an effective temperature of $\approx1100K$~\cite{Marois2008}, so the blackbody peak is at 4~\si{\um}, and a contrast of only $\sim$20--30~dB would be required. This case is even more favourable if the planet is surrounded by an accreting circumplanetary disk~\cite{Zhu2015}.

Whether through reflected light or emission of the exoplanet itself, nulling interferometry is another form of direct imaging, similar to coronagraph, that uses the star's own light to filter itself to extinction (a null) rather than using a physical block to omit the starlight from an exoplanet detection. Interferometry for exoplanet detection was proposed by Bracewell~\cite{Bracewell1978} in 1978. The idea was to use a space telescope system that could observe an entire star system and interfere light in a bulk optics configuration, separating out the light from the central star and the reflecting and emitting bodies, such as exoplanets or clouds of dust, around the star. The central star is said to be extinguished if the light from the mirrors interferes such that destructive interference occurs over the star. Due to the spatial separation of the star and any other orbiting body the phase difference in the multiple pathlengths will vary radially from the center of the telescope and the area around the central null will not be extinguished. The exoplanet is detectable as long as the exoplanet lies outside of a null fringe and is luminous enough to be detectable when compared to the nulled star.

Interferometry has been used in astronomy since Michelson~\cite{Michelson1891} measured the moons of Jupiter. But in modern astronomy, rather than using bulk optics to create Bracewell nulling interferometers, astronomers have turned to photonics. At present we are entering a renaissance of photonic systems replacing bulk optics and have begun to create devices never thought possible due to bulk optics limitations, like that of infrared kernel nulling instrument on the Very Large Telescope Interferometer~(VLTI) known as VIKiNG~\cite{Martinache2018}. This will roll onto future space missions that will become lower cost and more robust/stable due to massive optics on the order of kilograms being replaced with grams of monolithically integrated devices. Photonic systems are already proven on sky using Silica waveguide devices at  1550~nm~\cite{Norris2014,Errmann2015} and are now moving into the mid-infrared~(MIR) at 3.7~\si{\um}~\cite{Tepper2017} - the first in the astronomical L~band  (3.7~\si{\um}-4.2~\si{\um}).

Presently there are two approaches to creating nulling interferometers for the MIR. The first is the ultrafast laser direct write method of inscribing waveguides into a bulk Chalcogenide glass (ChG)~\cite{Arriola2014a,Gretzinger2015}. To create a nulling interferometer the direct write methods usually employ an evanescent coupler~\cite{Tepper2017}. Using Gallium Lanthanum Sulphide glass (GLS) waveguide losses as low as 0.25~dB/cm~\cite{Madden2013} have been demonstrated making it promising for astronomical applications. The second method uses physical vapour deposition techniques coupled with lithography, etching, and overcladding to create two dimensional fully etched waveguides that are buried in a chosen index cladding material. This method enables the use of multimode interference couplers (MMI) as 3~dB couplers~\cite{KenchingtonGoldsmith2016, KenchingtonGoldsmith2017a} which provide a potentially wide, fabricationally tolerant extinction bandwidth compared to conventional evanescent couplers. The two dimensional approach also provides a higher and tailorable index contrast between the core and cladding material allowing for much tighter bend radii  ($\approx200$~\si{\um} for the devices in this paper) allowing for chips to be more compact and with much smaller arm lengths than with direct write  this enhancing stability. 

The deposition and etch method is also in a key position to extend into the next generation of devices that not only null light (using a 3~dB coupler) but accounts for atmospheric disturbances. By using Kernal-Nulling~\cite{Martinache2018} a 4x4~coupler and a 3x6~coupler are all that are required to create an interferometer, with phase modulators between the couplers to account for any unwanted phase delay.

\section{Background}
The astronomical L~band is within an atmospheric window of the Earth which makes it a key area of research for Earth bound instruments. Photonic devices in the MIR often use materials such as Silicon on Sapphire~\cite{Li2011}, Silicon Germanium~\cite{Carletti2015}, Indium Phosphide~\cite{Smit2014}, Aluminium Gallium Arsenide~\cite{Ottaviano2016} and Lithium Niobate~\cite{Myers1996} due to Silica (SiO$_2$) being opaque in this region. ChG was chosen over the other materials because it offers a lower refractive index than all except Lithium Niobate, a widely tailorable refractive index contrast, has shown waveguide losses as low as 0.3~dB/cm~\cite{Yu2016} and is transparent beyond 10~\si{\um}, which is important for future work that will explore detections of cool blackbody objects with a temperature of 300~K and an ozone (O$_3$) absorption feature at 9.7~\si{\um}\cite{Seager2014}. 

Using standard deposition methods two ChGs were deposited: Germanium (Ge), Arsenic (As), Sulphur (S) as the cladding layers, and Ge, As, Selenium (Se) as the core layer. The layers were thermally evaporated from bulk glasses using a resistively heated crucible in a high vacuum chamber at $\approx10^{-6}$ Torr. The base wafers were Silicon (Si) with 5~\si{\um} of SiO$_2$ grown on top to optically segregate the two high index materials (Si and the ChG) stopping evanescent coupling to radiative modes in the Si. To produce a two dimensional device three layers were deposited: the bottom cladding - 3~\si{\um}, the core layer - 2~\si{\um}, and an additional top cladding layer - 1~\si{\um} - as an extra protective layer against the forthcoming plasma etching. 

A standard positive photoresist lithography process was used to define the waveguide pattern using a broadband exposure 1x~projection mask aligner. An Inductively Coupled Plasma~(ICP) etcher using CHF$_{3}$ as etch gas etched the upper cladding and core material that was not protected by the photoresist. The photoresist was then removed to prevent the polymer absorbing the MIR light. The wafer was put back into the deposition chamber for an overcladding deposition of the cladding glass to cover the core waveguide. The purpose of stacking the core layer between two cladding layers was to provide the 0.33~refractive index contrast around the entire waveguide.

\section{Throughput}
\begin{figure}[h]
\centering
	\includegraphics[width=0.75\columnwidth]{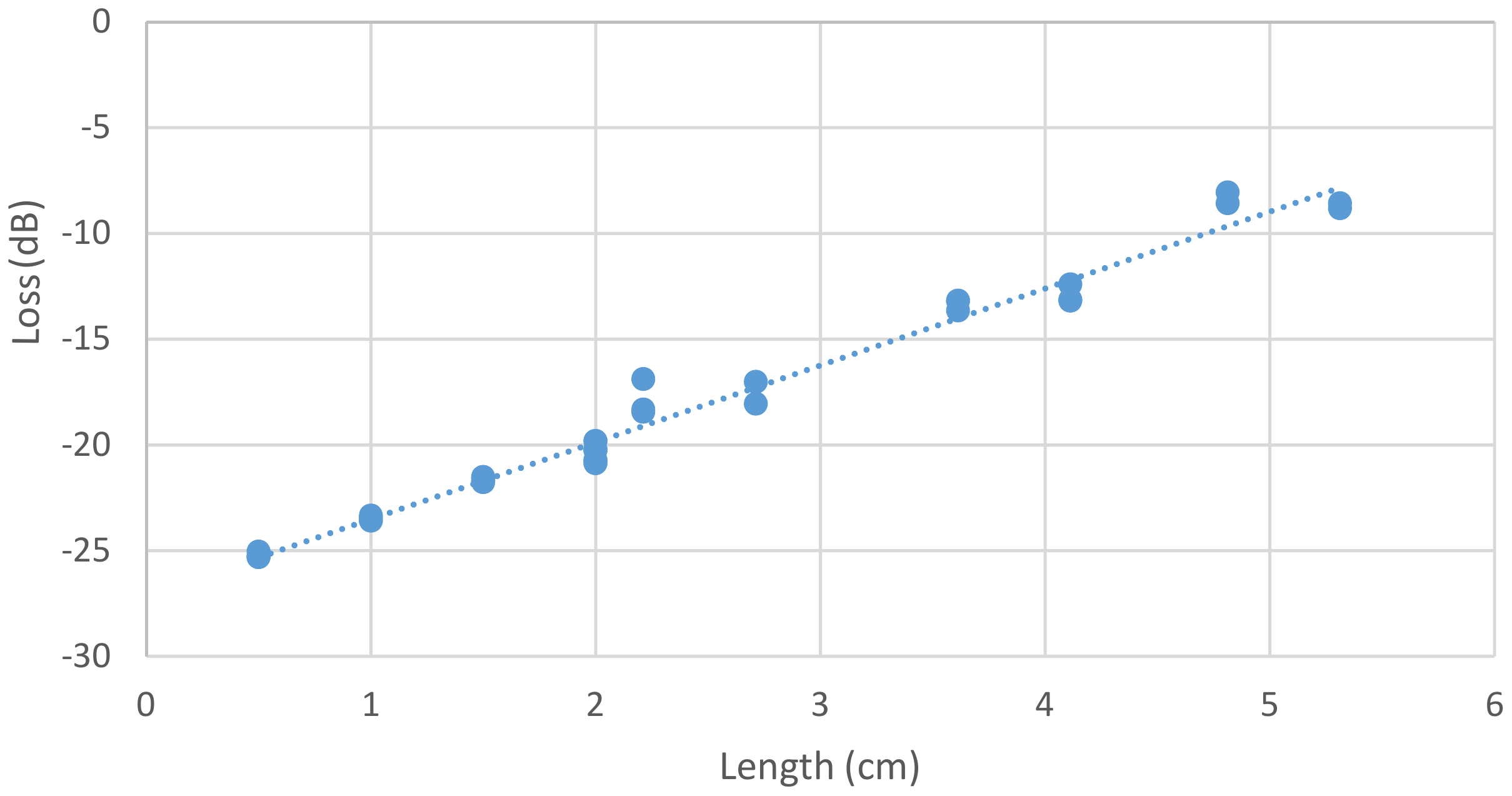}
  \caption{Cut back measurement for waveguide lengths of 0.5, 1 and 2~cm as well as for spirals of 2.3, 3.7, 4.9~\si{\cm} on a 1.5~cm long chip. All were of an overclad 2~\si{\um}~by~2~\si{\um} waveguide with the wavelength remaining at 4~\si{\um}.}
  \label{fig:4um}
\end{figure}

Based on previous results, low loss in the MIR was expected. Previous experiments had shown that the combination of Ge$_{11.5}$As$_{24}$S$_{64.5}$, with a refractive index of 2.279, and Ge$_{11.5}$As$_{24}$Se$_{64.5}$, with a refractive index of 2.609, has a loss of 0.3 dB/cm for a rib waveguide~\cite{Yu2016}. To reduce birefringence, however, fully etched waveguides are required for this work. Initial testing at 4~\si{\um} indicated that the loss for a fully etched waveguide was significantly higher than that previously measured in rib waveguides using this materials system. Figure~\ref{fig:4um} is a combination of loss measurements from straight waveguides and spirals and the linearity shows negligible bend losses (spiral bend radii were 200~\si{\um}).

\begin{figure}[h]
\centering
	\includegraphics[width=0.75\columnwidth]{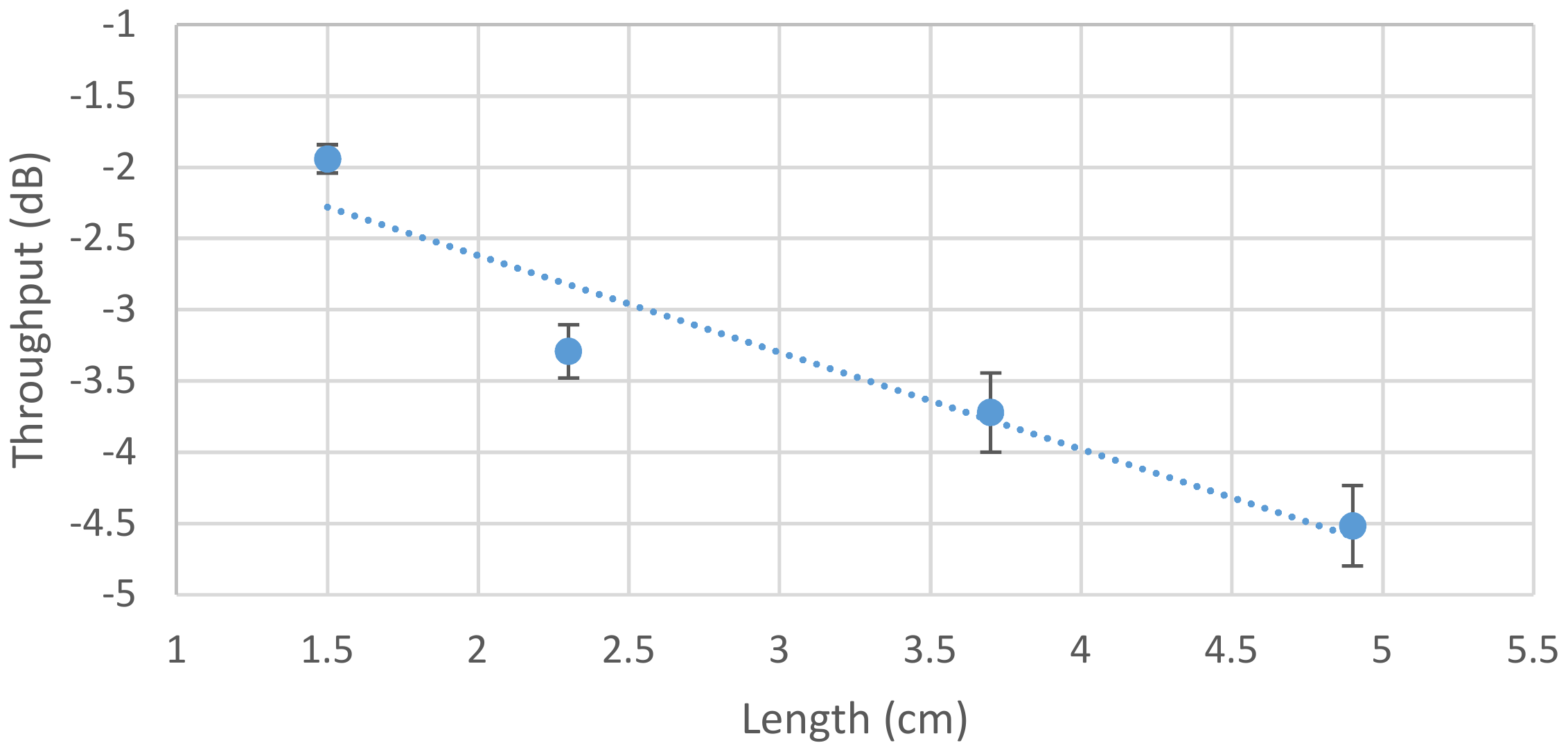}
  \caption{On-chip cutback, using spirals of lengths as in figure~\ref{fig:4um} for a wavelength of 1550~nm.}
  \label{fig:1550nm}
\end{figure}

The measurements were made on a variety of wafers and combined to give a loss measurement of $3.6\pm0.7$~dB/\si{\cm}. Waveguide loss was also measured at 1550~nm and, as shown in figure~\ref{fig:1550nm}, the loss was $0.7\pm0.3$~dB/cm. Given that the normally dominant loss source in high index contrast fully etched waveguides is sidewall roughness induced scattering (which scales roughly inverse quadratically with wavelength), then losses of 0.1~dB/cm would be expected at 4~\si{\um} if all absorptions are sufficiently attenuated.

\begin{figure}[h]
\centering
	\includegraphics[width=0.8\columnwidth]{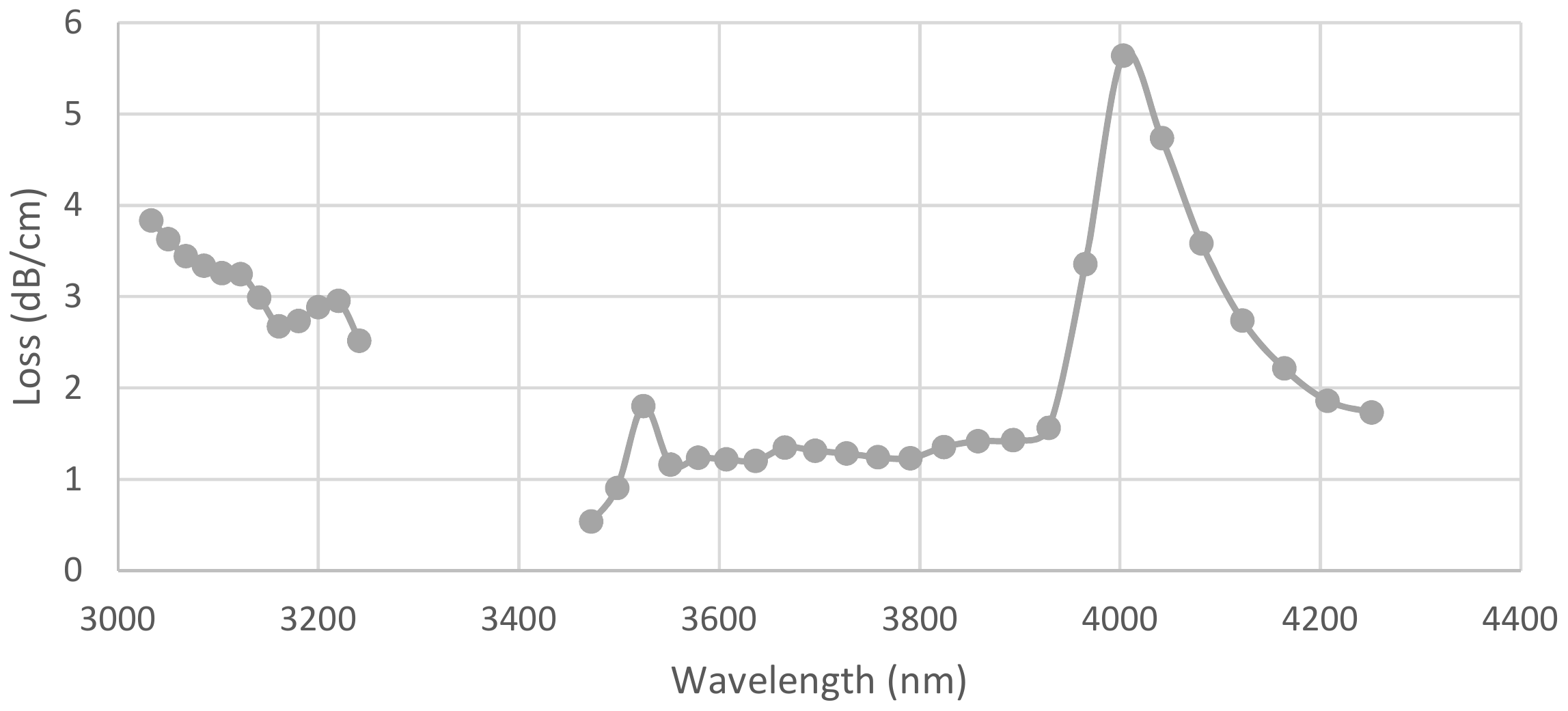}
  \caption{Loss measurement from 3000~nm to 4250~nm of a fully etched 3~\si{\um} by 2~\si{\um} air clad waveguide.}
  \label{fig:3umWG}
\end{figure}

The high loss at 4~\si{\um} was theorized to be due to a high concentration of S-H bonds in the cladding material. To test this, the loss was measured over a large wavelength range, in this case 3000~nm to 4250~nm as seen in figure~\ref{fig:3umWG}. 

The waveguides in figure~\ref{fig:3umWG} used air cladding and 3~\si{\um} wide (remaining 2~\si{\um} high) which means it is beyond the limit where the waveguide is strictly single mode when clad. However the asymmetric cladding causes it it remain single mode though at the expense of leaving the sidewalls vulnerable to atmospheric deposition of hydrocarbons and water with consequential loss increases~\cite{Ma2013}. Nonetheless it is clear that there was a large absorption feature at 4~\si{\um} and another absorption tail from 3000~nm likely due to water absorption. Using a finite difference waveguide simulation~\cite{Fallahkhair2008} it was estimated that the amount of light in the fully etched waveguide was 22\% compared to the rib waveguide at 8\% explaining why this peak was so much higher than in previous works~\cite{Ma2013}. Note the gap at 3200~nm to 3500~nm was due to low signal levels likely indicating significant C-H absorption from atmospherically chemisorbed species which is also likely the cause of the $\approx1$~dB/cm loss tail as previously observed~\cite{Ma2013}. It is clear from figure~\ref{fig:3umWG} that using glass with high concentrations of S is not ideal for astronomy.

\begin{figure}[h]
\centering
	\includegraphics[width=0.9\columnwidth]{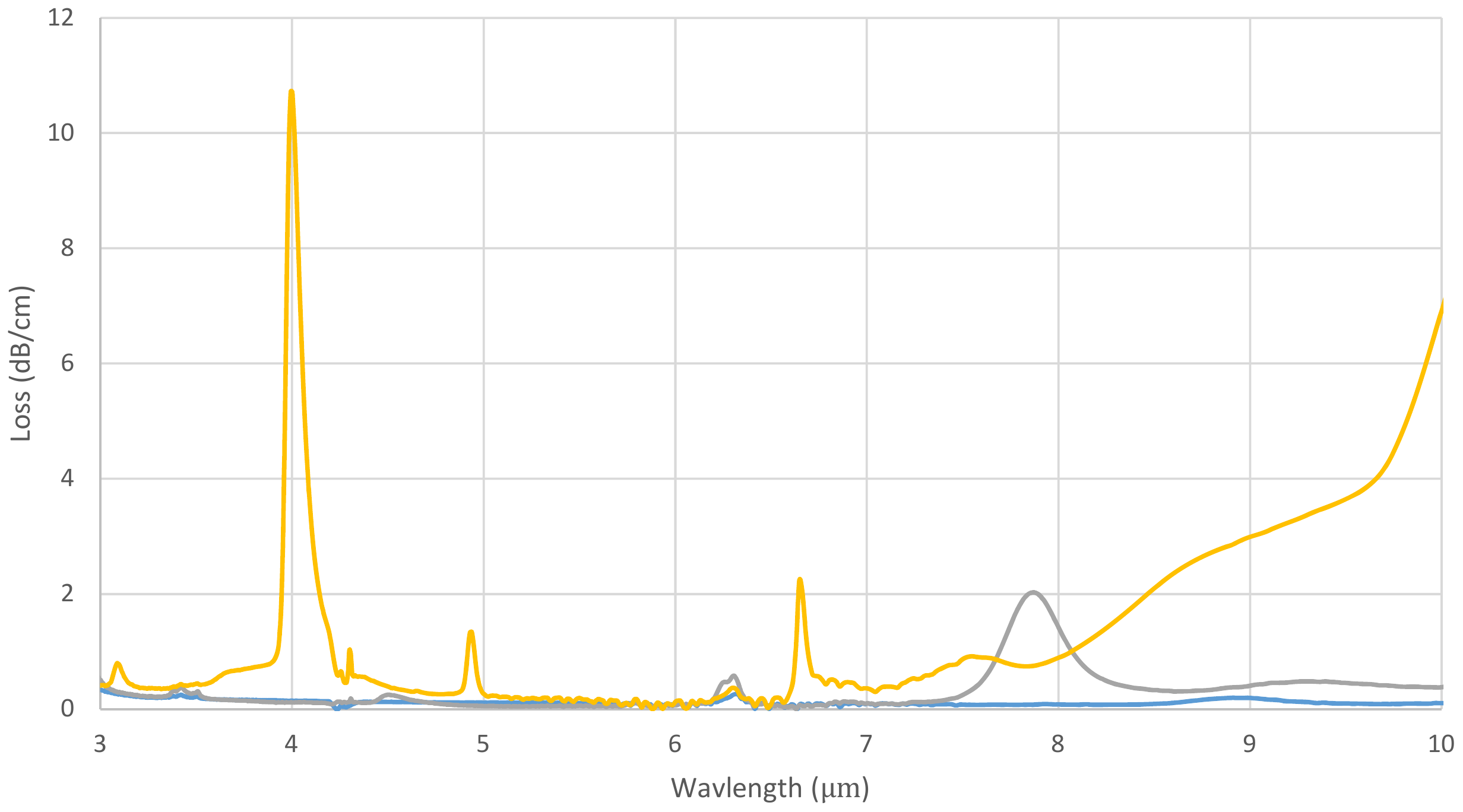}
  \caption{Bulk glass loss measurements for Ge$_{11.5}$As$_{24}$S$_{64.5}$ - yellow - Ge$_{20}$As$_{10}$Se$_{70}$ - grey - and As$_2$Se$_3$ - blue - normalized to a 1~cm thickness of bulk glass.}
  \label{fig:AllGlass}
\end{figure}

To reduce the loss further, a move to an all Selenide glass system is now being pursued. As$_2$Se$_3$ (also known as As$_{40}$Se$_{60}$) and Ge$_{20}$As$_{10}$Se$_{70}$  glasses have refractive indices of 2.835 as a replacement core layer and 2.502 as a replacement cladding layer respectively. They were chosen such that their index difference maintains the 0.33 difference as before, or as close to it as possible, so that the same photolithography masks can be used. Both glasses are Se based which eliminates any S-H absorption at 4~\si{\um}, with the Se-H absorption being both smaller and at 4.5~\si{\um} as shown in figure~\ref{fig:AllGlass} and so out of the band of interest. 

Figure ~\ref{fig:AllGlass} shows the spectra of the two new glasses compared to the Ge$_{11.5}$As$_{24}$S$_{64.5}$ glass. In the astronomical L-band region it is clear that the S glass has high loss - specifically at 4~\si{\um}  - whereas the new glasses have comparatively little loss. The Ge$_{20}$As$_{10}$Se$_{70}$ glass has a small absorption feature at 4.5~\si{\um} as expected from the Se-H absorption however the As$_2$Se$_3$ does not. Both glasses are also highly transmissive in the 10~\si{\um} region compared to the S glass; important for future extension and looking for O$_3$ absorption.

\section{Imbalance}
\begin{figure}[h]
	\centering
	\begin{subfigure}{0.49\textwidth}
		\includegraphics[width=1.00\textwidth]{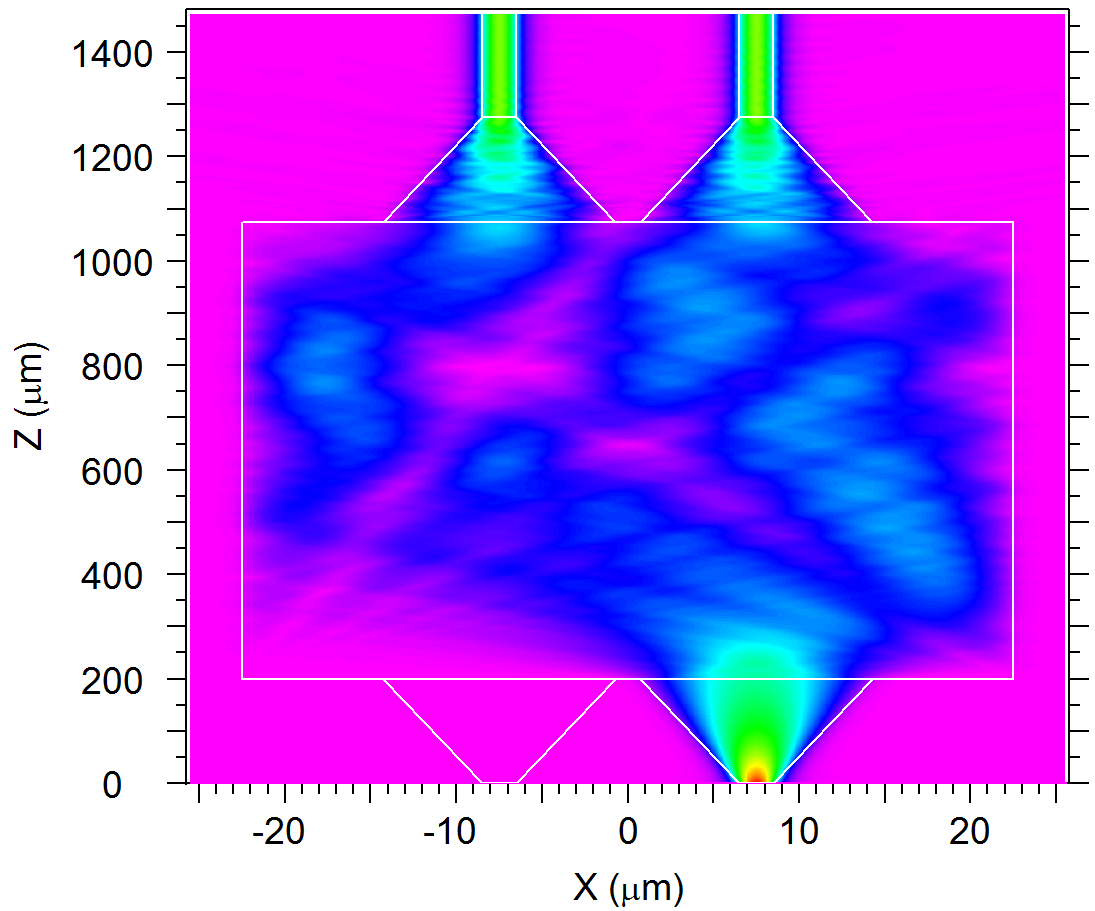}
        \caption{3~dB coupler.}
        \label{fig:MMI}
     \end{subfigure}
\hfill
	\begin{subfigure}{0.49\textwidth}
		\includegraphics[width=1.00\textwidth]{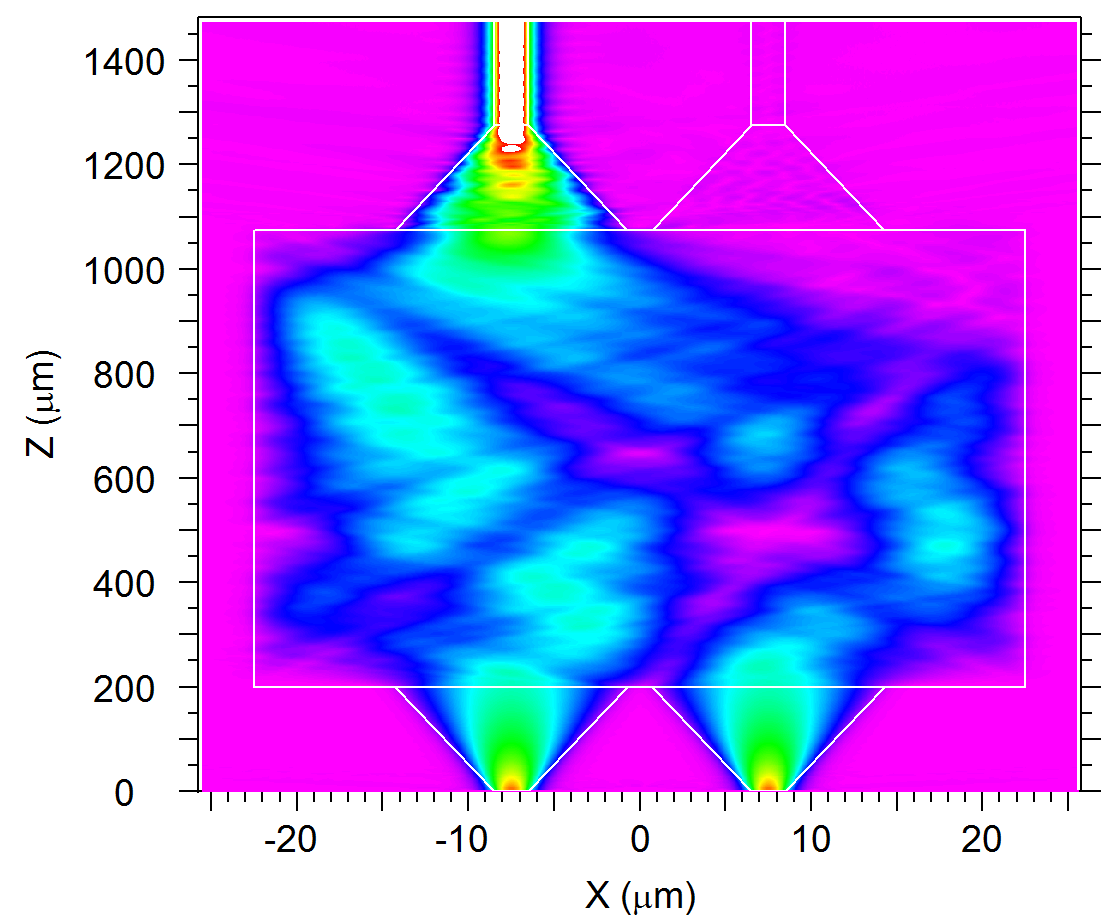}
        \caption{Beam combiner.}
        \label{fig:Combiner}
     \end{subfigure}
     \caption{An MMI of width 45~\si{\um}, length 875~\si{\um} and taper width of 13.5~\si{um}~\cite{KenchingtonGoldsmith2017a}. The light is injected from the bottom to the top.}
     \label{fig:Simulations}
\end{figure}

In order to split exoplanet light from stellar light using interferometry, a 3~dB~coupler or network of such is required. The purpose of the coupler is to allow the self interference as described by Bracewell~\cite{Bracewell1978} of the star such that the stellar light is directed into a single output port. This is shown in figure~\ref{fig:Combiner} with the inputs at the bottom. For the MMI to produce a combination function with high extinction in the other port as shown in figure~\ref{fig:Combiner} the phase between the inputs must be $90^{\circ}$ ~\cite{Soldano1995,KenchingtonGoldsmith2017a}. A 3~dB~coupler works the same way, as shown in figure~\ref{fig:MMI}, but in reverse: taking in one input and splitting it evenly. Hence, testing the MMI as a 3~dB~coupler is indicative of using it as a combiner forming a null as in the right port in figure~\ref{fig:Combiner}.

\begin{figure}[h]
\centering
	\includegraphics[width=0.7\columnwidth]{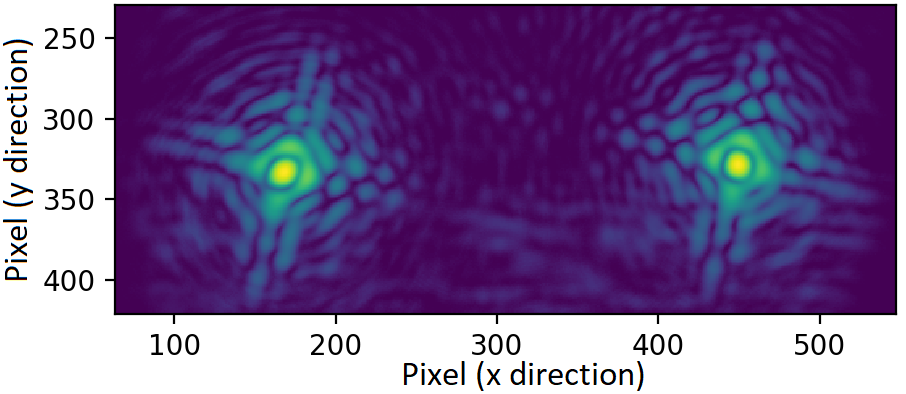}
  \caption{Lock-in image from output waveguides of an MMI, as shown in figure~\ref{fig:MMI}. Colour scale is an arcsinh stretch, with scattered light background RMS of 2$\times 10^{-3}$ in
  between the two bright outputs. }
  \label{fig:ImbalanceImages}
\end{figure}

The initial test of a 3~dB coupler is in its ability to split light evenly. As previously published~\cite{KenchingtonGoldsmith2016} a 2x2~MMI can be manufactured with a splitting ratio of almost exactly 50:50 over a large bandwidth. However at the time of that publication the measurement accuracy was unclear. Replacing the pyroelectric detectors~\cite{KenchingtonGoldsmith2016} with an InSb camera (Xenics infrared solutions - Onca MWIR) and using lock in detection, the background noise was reduced, much of the scattered light filtered out and the integration of the signal boosted producing images like that in figure~\ref{fig:ImbalanceImages}.

\begin{figure}[h]
	\centering
	\begin{subfigure}{0.49\textwidth}
	\includegraphics[width=\columnwidth]{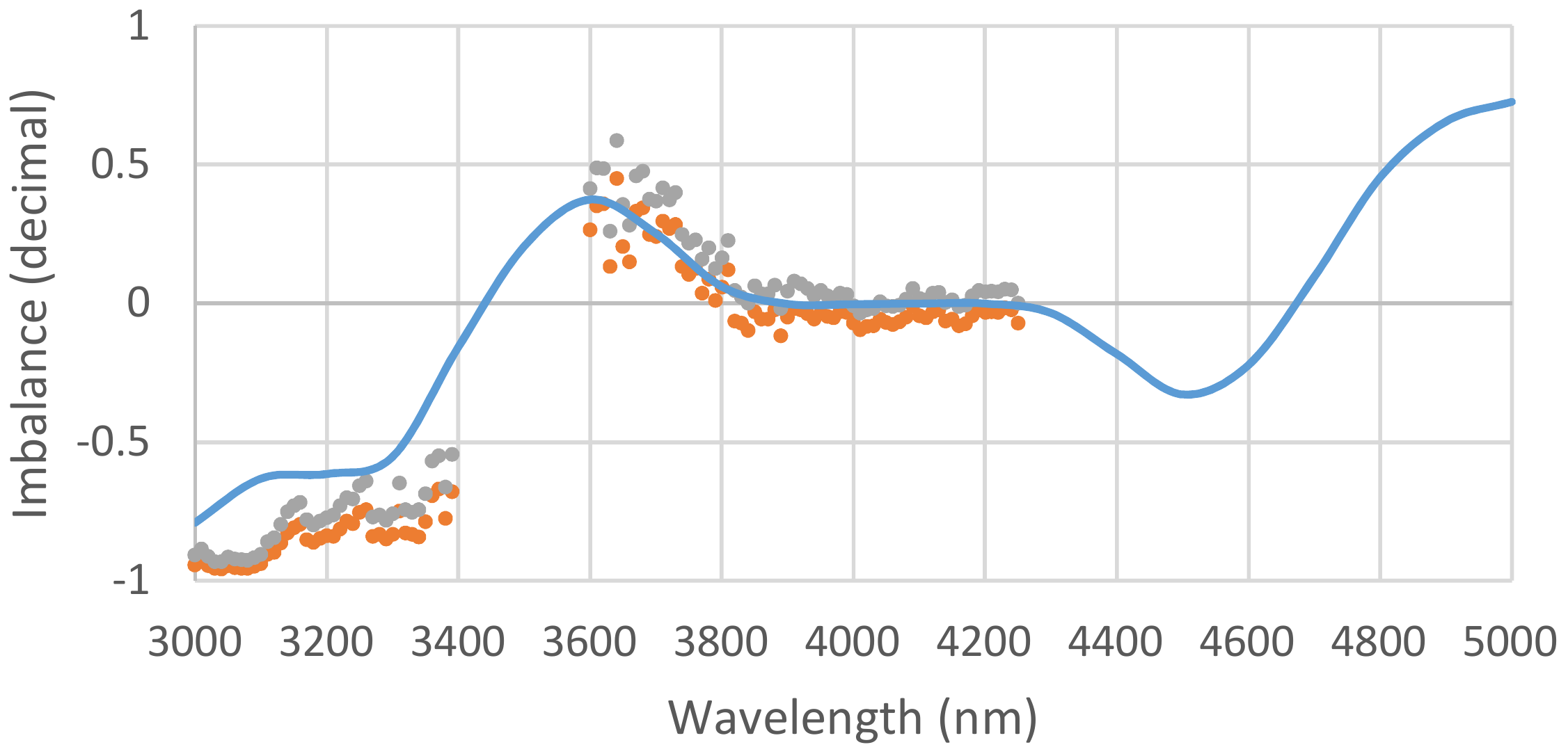}
  \caption{MMI input-output taper width of 8~\si{\um}.}
  \label{fig:8umTaper}
     \end{subfigure}
\hfill
	\begin{subfigure}{0.49\textwidth}
	\includegraphics[width=\columnwidth]{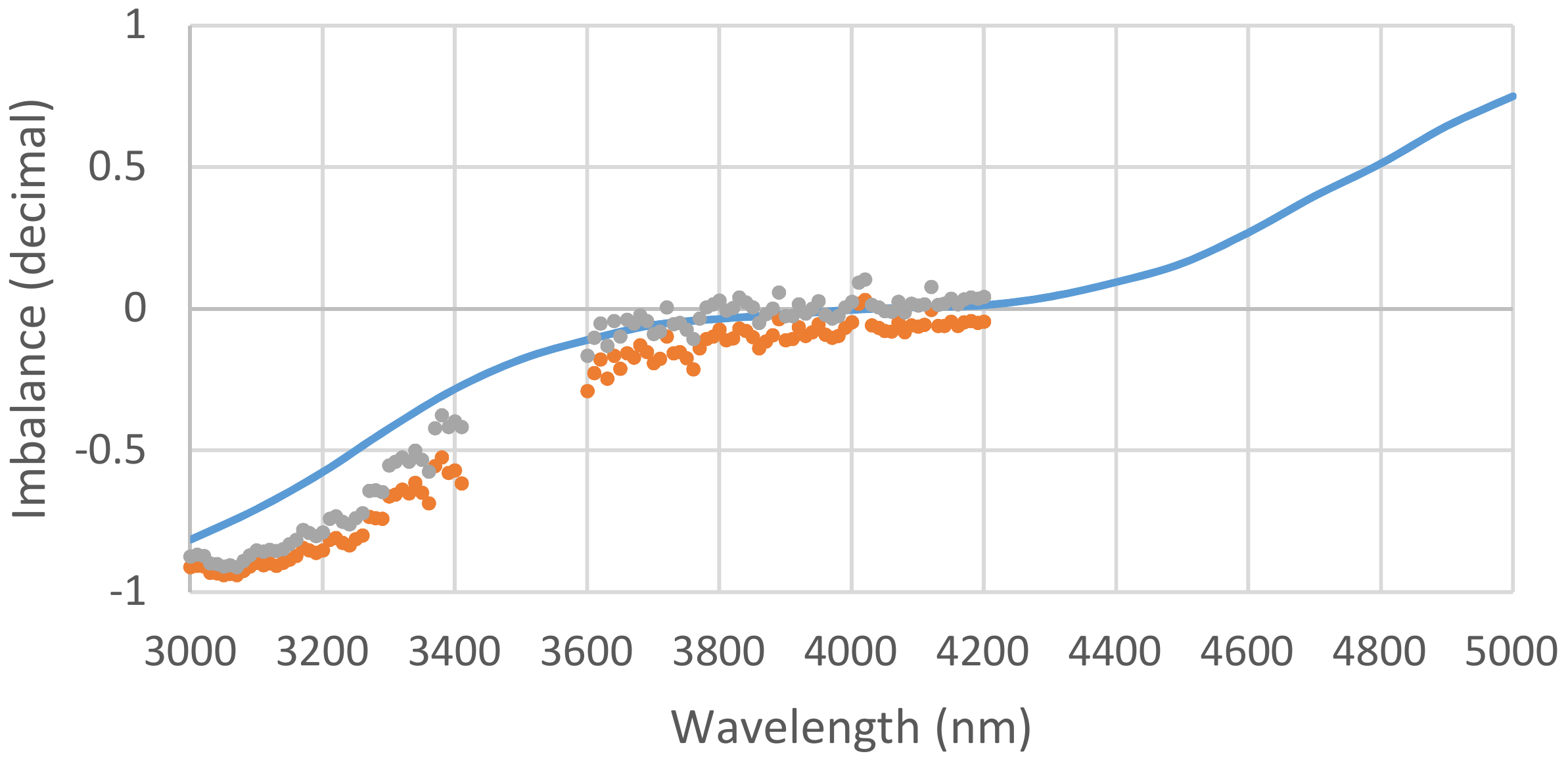}
  \caption{MMI input-output taper width of 13.5~\si{\um}.}
  \label{fig:13umTaper}
  \end{subfigure}
  \caption{The simulated (in blue) and experimental (upper bound in grey, lower bound in orange) imbalance of an MMI as shown in figure~\ref{fig:MMI}.}
  \label{fig:experimentalImbalance}
\end{figure}

Using the maximum value of the two localized intensity maxima, the bright spots in figure~\ref{fig:ImbalanceImages}, the imbalance over the L-band is calculated:
\begin{equation}
\text{Imbalance (decimal)} = \frac{\text{bar port}-\text{cross port}}{\text{bar port}+\text{cross port}}.
\label{eq:imbalance}
\end{equation}
Comparing simulation to the experimental results in figure~\ref{fig:experimentalImbalance} shows a good fit between the two, and within error bounds the MMIs have an imbalance better than 0.1 across significant bandwidth. 

Note here that the figures have an upper and lower bound for the experimental data. This was an attempt to put bounds on the effects of scattered and uncoupled input light on the output fields. These noise terms interfere coherently drastically increasing the effect of scattered light on the data. The two sets of data represent the values calculated by adding and subtracting the average scattered light between the two ports, as shown in figure~\ref{fig:ImbalanceImages}. This was done in electric field terms and 
\begin{equation}
\text{Corrected Intensity}_{\pm}=(\sqrt{\text{Measured intensity}}\pm\sqrt{\text{Scattered Light}})^{2}.
\label{eq:ElecFieldCalc}
\end{equation}
Looking again at figure~\ref{fig:ImbalanceImages} there is little scattered light visible to the naked eye but the calculated average scattered light intensity of 2 compared to a peak of 7900, using equation~\eqref{eq:ElecFieldCalc}, provides a maximum or minimum value of 7600 and 8100 or a $\approx3.5\%$ difference compared to the much smaller 0.03\% if using intensity addition and subtraction. On that note a scattered light intensity value of 2 is only 36~dB lower than a signal of 8000 and thus significantly restricts the measurable depth of a potential null.

From previous work~\cite{KenchingtonGoldsmith2017b} an imbalance of 0.0018 is analogous to 60~dB of extinction in an equivalent nulling device or Mach-Zehnder interferometer (MZI). Due to the scattered light the coupling ratio could not be measured to this level of accuracy and so the absolute performance of the MMI coupler could not be verified. Using the same calculations an imbalance of 0.1 provides an extinction value of 20~dB. Thus figure~\ref{fig:8umTaper} shows that a null of 20~dB should be achievable over a 450~nm bandwidth as the upper and lower bound from 3800~nm to 4250~nm are within an imbalance of $\pm0.1$.

The ideal MMI, with the highest throughput, has a taper width of 13.5~\si{\um}~\cite{KenchingtonGoldsmith2017a}. From figure~\ref{fig:13umTaper} this MMI has a limited bandwidth where the extinction would be maximized. This design passes through zero imbalance at a single point rather than making multiple passes through zero, as the 8~\si{\um} taper width does, though this is not clearly visible in figure~\ref{fig:8umTaper}. Both sets of data in figure~\ref{fig:ImbalanceImages} follow the expected trends as their simulation predicts from 3600~nm, but both deviate at shorter wavelengths. The reasons for this are unclear at present but it is not a major concern as it is well outside the operational bandwidth.

\section{Null}
By placing two MMIs in series, placing figure~\ref{fig:MMI} below figure~\ref{fig:Combiner}, the MMI was tested as a nulling interferometer in a Mach-Zehnder configuration. Numerical simulations based on the 8~\si{\um} MMI taper width predict an extinction greater than 60~dB~\cite{KenchingtonGoldsmith2017b} however this does not account for any scattered light or pathlength difference between the two arms as the light travels between the two MMIs.

\begin{figure}[h]
\centering
	\includegraphics[width=0.9\columnwidth]{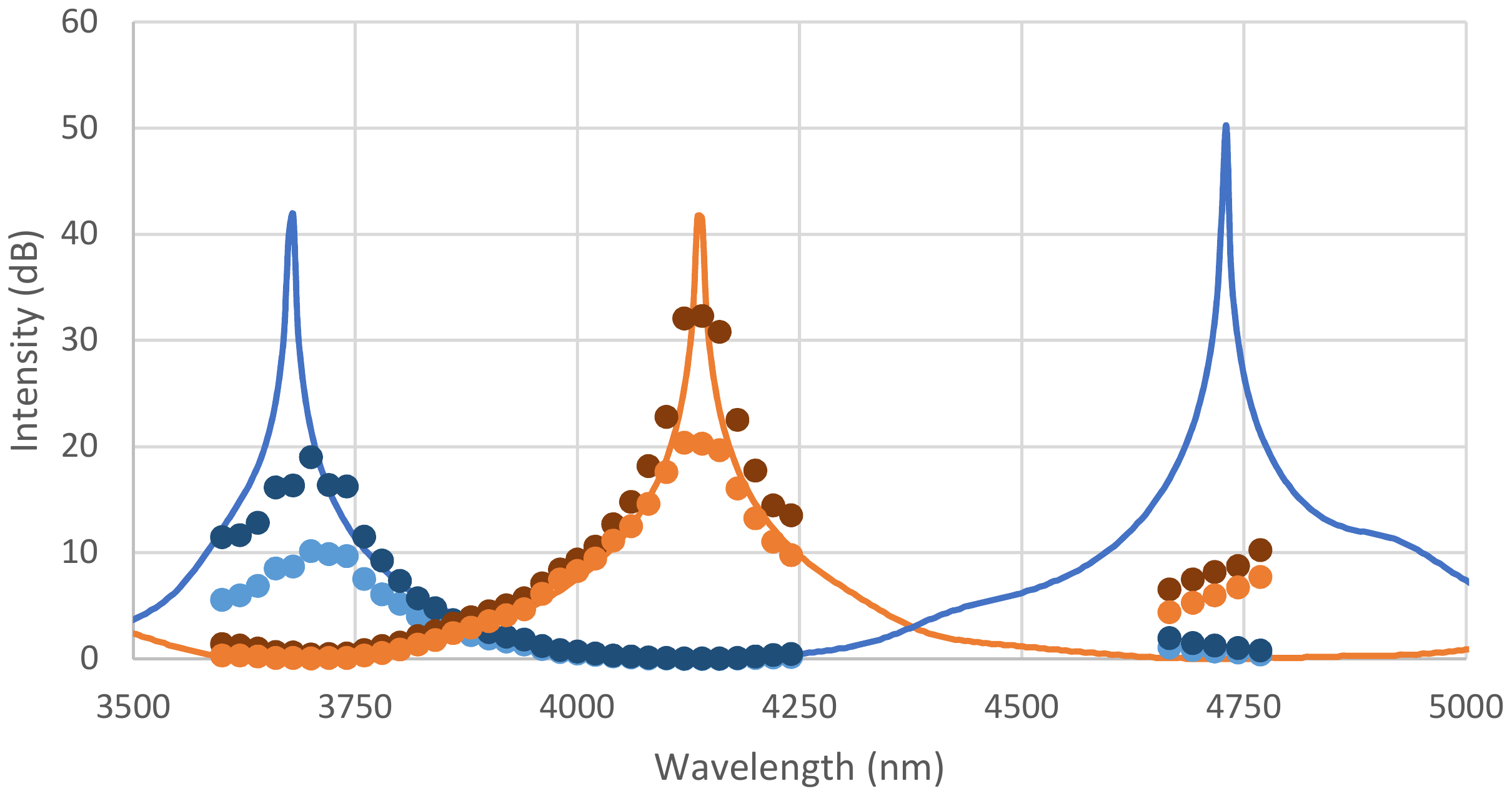}
  \caption{The individual intensities of the MZI arms, with upper and lower bonds of arm one in dark orange and orange and for arm two in dark blue and blue. The orange solid line is a theoretical match based on the patherlength difference of the two arms of the MZI for arm one (from equation~\eqref{eq:MZITheory} this is $-10Log_{10}(A1-B)$), and the blue solid line is for arm two ($-10\log_{10}(A2+B)$). The $\Delta OPL$ was set to 16.55~\si{\um}.}
  \label{fig:ArmsOfFirstExtinction}
\end{figure}

Using basic theoretical model, in equation~\eqref{eq:MZITheory}, of the electric field through the MZI the effect of an optical pathlength difference ($\Delta OPL$) was compared to an experimental measurement of a device constructed from two MMIs, with taper widths of \textbf{8}~\si{\um}, and a 14~mm long straight waveguide interconnect them. 

\begin{equation}
\begin{split}
A1 &=\cos(C1)^2\cos(C2)^2 + \sin(C1)^2\sin(C2)^2\\
A2 &=\cos(C1)^2sin(C2)^2 +\sin(C1)^2\cos(C2)^2\\
B &= 2\cos(C1)\cos(C2)\sin(C1)\sin(C2)\cos(2\pi\Delta OPL/\lambda)\\
Extinction &= 10\log_{10}(\frac{A2 + B}{A1 - B})\\
&\text{Where C1 and C2 are wavelength dependent coupling ratios generate by Rsoft.}
\end{split}
\label{eq:MZITheory}
\end{equation}

From figure~\ref{fig:ArmsOfFirstExtinction} it is clear that the model, using an $\Delta OPL$ of 16.55~\si{\um}, in equation~\eqref{eq:MZITheory} fits the shape of what was experimentally measured. The intensity at the peaks is an exception to this, due to scattered light limiting the ability to detect further than $\sim$30~dB, and at wavelengths beyond 4500~nm. It is unclear why the theory does not match the larger wavelengths and more experiments are required to test the theory further.

\begin{figure}[h]
\centering
	\includegraphics[width=0.9\columnwidth]{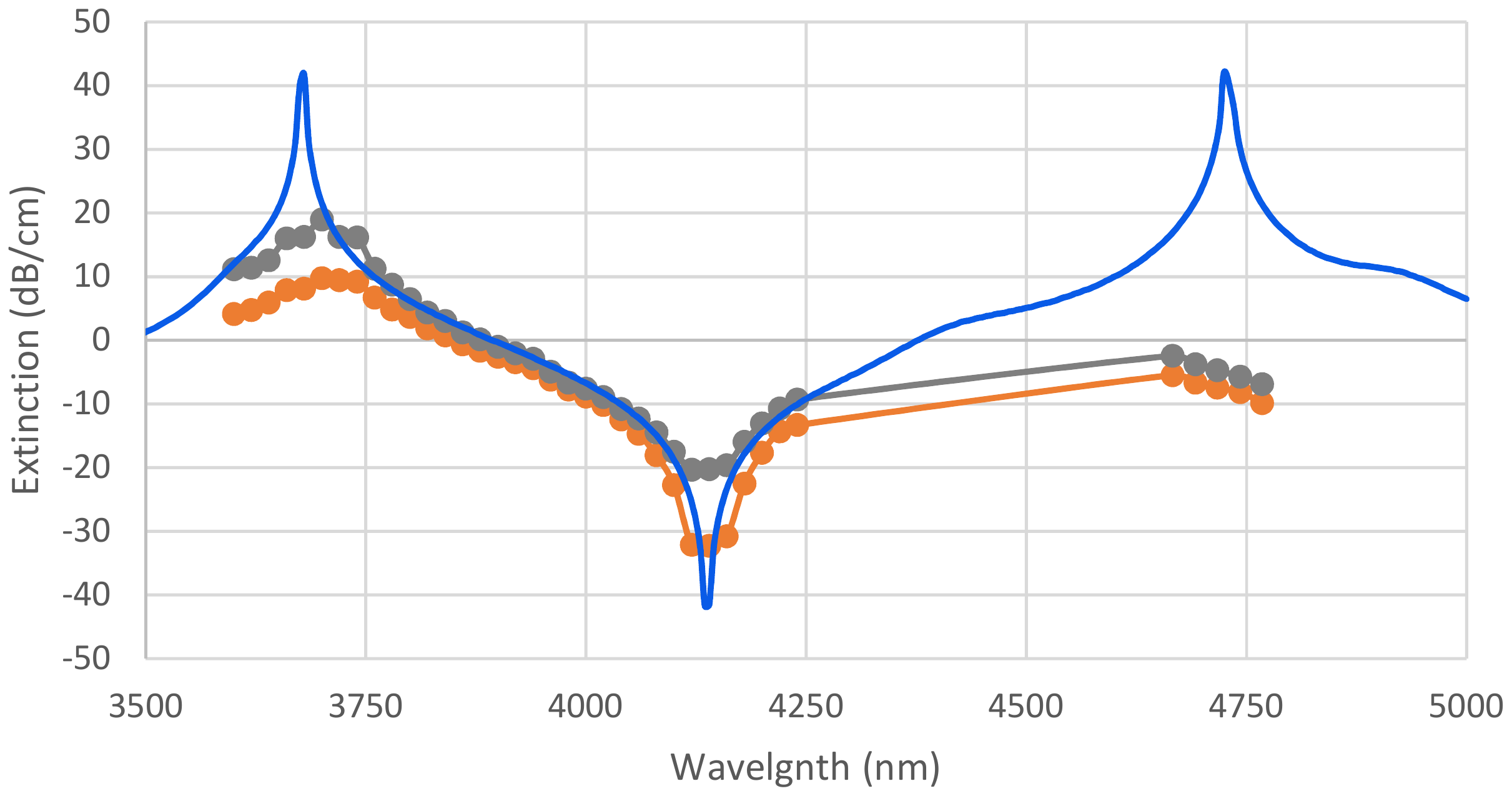}
  \caption{The first measurement of extinction of two MMI in sequence over a wavelength of 3600~nm to 4750~nm with upper and lower bonds as per figure~\ref{fig:experimentalImbalance}. The blue solid line is a theoretical match based on the patherlength difference of the two arms of the MZI as per equation~\eqref{eq:MZITheory}.}
  \label{fig:FirstExtinction}
\end{figure}

The extinction calculation thus from figure~\ref{fig:ArmsOfFirstExtinction} is represented as figure~\ref{fig:FirstExtinction}. This clearly shows a null down to the measurement limited predicted depth of -20~dB (following the upper limit grey curve of figure~\ref{fig:FirstExtinction}, at 4150~nm With a reflected 20~dB peak at 3600~nm, however the null should be at 4000~nm~\cite{KenchingtonGoldsmith2017b} likely shifted due to the $\Delta OPL$.

Figures~\ref{fig:ArmsOfFirstExtinction}~and~\ref{fig:FirstExtinction} illustrate the effect a difference in the optical path lengths between the two MMI ports can have despite being designed to be equal. The reason behind such a drastic change is the change in phase of the light as it enters the second MMI. From the two extinctions in figure~\ref{fig:FirstExtinction}, at 3600~nm and 4150~nm, the free spectral range (FSR) is calculated to be $1100$~nm.
Using
\begin{equation}
FSR = \frac{\lambda^2}{\Delta n\times Length}=\frac{\lambda^2}{\Delta OPL}
\label{eq:FSR}
\end{equation}
the $\Delta OPL$ is found to be $\approx$15000~nm (15~\si{\um}), which is similar to the 16.55~\si{\um} used in the theory. From the known arm length this implies an effective refractive index difference of $1\times10^{-3}$ between the arms is sufficient to achieve such a wavelength dependence. Using Rsoft FEMSim Finite Element Simulation it was found that a 25~nm difference in the width of the waveguides is enough to achieve this change in effective refractive index. Given the critical dimension tolerance on the photomask of $\pm$100~nm, and the specifications of the mask writer, coupled with the use of 1x~projection lithography, this level of dimensional variation is not surprising. It also points to the need to keep the devices as compact as possible and also to the need to employ measures to minimise the effects of such dimensional variations. 

One approach that can be solve this issue is to use thermo-optic phase shifters formed with resistive heaters above the waveguides. Due to ChG being a good thermal insulator there is little worry that the heat will spread between the arms. Thus the $\Delta OPL$ will be addressed in future work.

\section{Summary}
For the mid infrared of the electromagnetic spectrum, photonic chips are undergoing a renaissance with direct write fabrication of nulling interferometers being produced for the first time in a high throughput glass. In tandem the two dimensional, fully etched, fabrication is producing couplers with great accuracy that fit simulations. Scattered light is still a factor in the measurement accuracy of these couplers where future work will need different architectures and approaches to cladding mode stripping to sufficiently suppress the small amounts of light remaining and interfering with the light in the output ports. 

Waveguide loss is still a work in progress but a solvable problem based on prior results which promise losses below 0.5~dB/cm across the MIR band up to $>10$~\si{\um}. It also needs to be appreciated that device sizes are very small with the high index technology: the MMIs are 2~mm long and the interconnects can be manufactured to far smaller lengths due to the small bend radius possible with this technology. This enhances stability from thermal and mechanical effects, and reduces the degree of active trimming required as well as reducing the total insertion loss.

This paper is the first, to our knowledge, to show the results of a nulling interferometer in the astronomical L~band. It was shown that a measurement limited 20~dB null can be produced however, due to the optical pathlength difference, the null is not as deep nor as wide as predicted. Future work will correct the effective refractive index difference between interconnections of the MZI to form a null at 4~\si{\um} and suppress scattered light to enable much deeper nulls to be generated and measured. 

Once this is accomplished, further work will look past simply nulling out starlight but also accounting for atmospheric disturbances in potential exoplanet detections. By adapting the MMI to accept four inputs the nuller could become a Kernal-nuller, with the Cr heaters overlaid on the three output waveguides to correct any phase change during transit into the 3x6~sensing~coupler.

\acknowledgments 
 
This research was supported by the Australian Research Council (ARC) Centre of Excellence for Ultrahigh bandwidth Devices for Optic Systems (CUDOS) project CE110001018.

\bibliography{report} 
\bibliographystyle{spiebib} 

\end{document}